\documentclass[onecolumn,11pt]{article}
\usepackage[top=1in, bottom=1in, left=1in, right=1in]{geometry}
\usepackage{amsmath,amsfonts,amscd,amssymb}
\usepackage{graphicx}
\graphicspath{{figures/}}
\usepackage{url}
\usepackage{xcolor}
\usepackage{caption}
\usepackage{mathtools}
\usepackage{setspace}
\usepackage{cleveref}
\usepackage{authblk}

\usepackage{caption}
\usepackage{subcaption}
\usepackage{bm}

\usepackage{tikz}
\usetikzlibrary{shapes,arrows,calc,positioning,fit}
\usepackage[numbers,sort&compress]{natbib}

\usepackage{listings}
\usepackage{xcolor}

\definecolor{maroon}{cmyk}{0, 0.87, 0.68, 0.32}
\definecolor{halfgray}{gray}{0.55}
\definecolor{ipython_frame}{RGB}{207, 207, 207}
\definecolor{ipython_bg}{RGB}{247, 247, 247}
\definecolor{ipython_red}{RGB}{186, 33, 33}
\definecolor{ipython_green}{RGB}{0, 128, 0}
\definecolor{ipython_cyan}{RGB}{64, 128, 128}
\definecolor{ipython_purple}{RGB}{170, 34, 255}

\lstdefinelanguage{iPython}{
    morekeywords={access,and,break,class,continue,def,del,elif,else,except,exec,finally,for,from,global,if,import,in,is,lambda,not,or,pass,print,raise,return,try,while,True,False,as},%
    morekeywords=[2]{abs,all,any,basestring,bin,bool,bytearray,callable,chr,classmethod,cmp,compile,complex,delattr,dict,dir,divmod,enumerate,eval,execfile,file,filter,float,format,frozenset,getattr,globals,hasattr,hash,help,hex,id,input,int,isinstance,issubclass,iter,len,list,locals,long,map,max,memoryview,min,next,object,oct,open,ord,pow,property,range,raw_input,reduce,reload,repr,reversed,round,set,setattr,slice,sorted,staticmethod,str,sum,super,tuple,type,unichr,unicode,vars,xrange,zip,apply,buffer,coerce,intern},%
    sensitive=true,%
    morecomment=[l]\#,%
    morestring=[b]',%
    morestring=[b]",%
    morestring=[s]{'''}{'''},
    morestring=[s]{"""}{"""},
    morestring=[s]{r'}{'},
    morestring=[s]{r"}{"},%
    morestring=[s]{r'''}{'''},%
    morestring=[s]{r"""}{"""},%
    morestring=[s]{u'}{'},
    morestring=[s]{u"}{"},%
    morestring=[s]{u'''}{'''},%
    morestring=[s]{u"""}{"""},%
    literate=
    {á}{{\'a}}1 {é}{{\'e}}1 {í}{{\'i}}1 {ó}{{\'o}}1 {ú}{{\'u}}1
    {Á}{{\'A}}1 {É}{{\'E}}1 {Í}{{\'I}}1 {Ó}{{\'O}}1 {Ú}{{\'U}}1
    {à}{{\`a}}1 {è}{{\`e}}1 {ì}{{\`i}}1 {ò}{{\`o}}1 {ù}{{\`u}}1
    {À}{{\`A}}1 {È}{{\'E}}1 {Ì}{{\`I}}1 {Ò}{{\`O}}1 {Ù}{{\`U}}1
    {ä}{{\"a}}1 {ë}{{\"e}}1 {ï}{{\"i}}1 {ö}{{\"o}}1 {ü}{{\"u}}1
    {Ä}{{\"A}}1 {Ë}{{\"E}}1 {Ï}{{\"I}}1 {Ö}{{\"O}}1 {Ü}{{\"U}}1
    {â}{{\^a}}1 {ê}{{\^e}}1 {î}{{\^i}}1 {ô}{{\^o}}1 {û}{{\^u}}1
    {Â}{{\^A}}1 {Ê}{{\^E}}1 {Î}{{\^I}}1 {Ô}{{\^O}}1 {Û}{{\^U}}1
    {œ}{{\oe}}1 {Œ}{{\OE}}1 {æ}{{\ae}}1 {Æ}{{\AE}}1 {ß}{{\ss}}1
    {ç}{{\c c}}1 {Ç}{{\c C}}1 {ø}{{\o}}1 {å}{{\r a}}1 {Å}{{\r A}}1
    {€}{{\EUR}}1 {£}{{\pounds}}1,
    literate=
    *{-}{{{\color{ipython_purple}-}}}1
     {?}{{{\color{ipython_purple}?}}}1
     {^}{{{\color{ipython_purple}\^{}}}}1
     {=}{{{\color{ipython_purple}=}}}1
     {+}{{{\color{ipython_purple}+}}}1
     {*}{{{\color{ipython_purple}$^\ast$}}}1
     {/}{{{\color{ipython_purple}/}}}1
     {>}{{{\color{ipython_purple}$>$}}}1
     {<}{{{\color{ipython_purple}$<$}}}1
     {@}{{{\color{ipython_purple}@}}}1
     {+=}{{{+=}}}1
     {-=}{{{-=}}}1
     {*=}{{{$^\ast$=}}}1
     {/=}{{{/=}}}1,
    identifierstyle=\color{black}\ttfamily,
    commentstyle=\color{ipython_cyan}\ttfamily,
    stringstyle=\color{ipython_red}\ttfamily,
    keepspaces=true,
    showspaces=false,
    showstringspaces=false,
    rulecolor=\color{ipython_frame},
    backgroundcolor=\color{ipython_bg},
    basicstyle=\normalsize,
    keywordstyle=\color{ipython_green}\ttfamily,
}

\definecolor{mygray}{gray}{0.95}
\definecolor{codegreen}{rgb}{0,0.6,0}
\definecolor{codegray}{rgb}{0.5,0.5,0.5}
\definecolor{codepurple}{rgb}{0.58,0,0.82}
\definecolor{backcolour}{rgb}{0.95,0.95,0.92}

\lstdefinestyle{mystyle}{
    backgroundcolor=\color{mygray},   
    commentstyle=\color{codegreen},
    keywordstyle=\color{magenta},
    numberstyle=\tiny\color{codegray},
    stringstyle=\color{codepurple},
    basicstyle=\ttfamily\footnotesize,
    breakatwhitespace=false,         
    breaklines=true,                 
    captionpos=b,                    
    keepspaces=true,                 
    showspaces=false,                
    showstringspaces=false,
    showtabs=false,                  
    tabsize=2
}

\usepackage[bottom,flushmargin,hang,multiple]{footmisc}
\usepackage{lipsum}
\newcommand\blfootnote[1]{%
  \begingroup
  \renewcommand\thefootnote{}\footnote{#1}%
  \addtocounter{footnote}{-1}%
  \endgroup
}

\definecolor{header1}{cmyk}{0,0,0,1}

\newcommand{\bF}{\mathbf{F}}
\newcommand{\bA}{\mathbf{A}}
\newcommand{\bB}{\mathbf{B}}
\newcommand{\flow}{\mathbf{F}}
\newcommand{\bK}{\mathbf{K}}
\newcommand{\bC}{\mathbf{C}}

\newcommand{\bW}{\mathbf{W}}
\newcommand{\bX}{\mathbf{X}}

\newcommand{\bdyn}{\mathbf{f}}
\newcommand{\bg}{\mathbf{g}}
\newcommand{\bx}{\mathbf{x}}
\newcommand{\bu}{\mathbf{u}}
\newcommand{\bv}{\mathbf{v}}

\newcommand{\bz}{\mathbf{z}}

\newcommand{\bTheta}{\boldsymbol{\Theta}}
\newcommand{\bPhi}{\boldsymbol{\Phi}}
\newcommand{\bPsi}{\boldsymbol{\Psi}}
\newcommand{\bLambda}{\boldsymbol{\Lambda}}
\newcommand{\bxi}{\boldsymbol{\xi}}
\newcommand{\ddt}{\frac{\text{d}}{\text{d}t}}

\newcommand{\Koop}{\mathcal{K}}
\newcommand{\LieOp}{\mathcal{L}}

\tikzstyle{block} = [rectangle, draw, fill=blue!20, 
    text width=4.5em, text centered, rounded corners, minimum height=4em]
\tikzstyle{line} = [draw, -latex']

\title{\vspace{-.65in}{\fontsize{16}{16}\selectfont \textbf{PyKoopman: A Python Package for Data-Driven Approximation of the Koopman Operator}}\vspace{-.15in}}

\author[1,3]{\normalsize{Shaowu Pan}*}
\author[2]{Eurika Kaiser}
\author[1]{Brian M. de Silva}
\author[1]{\\J. Nathan Kutz}
\author[2]{Steven L. Brunton \vspace{-0.15in}}
\affil[1]{\footnotesize Department of Applied Mathematics, University of Washington, Seattle, WA 98195, United States}
\affil[2]{Department of Mechanical Engineering, University of Washington, Seattle, WA 98195, United States}
\affil[3]{Department of Mechanical, Aerospace, and Nuclear Engineering, Rensselaer Polytechnic Institute, \protect\\ Troy, NY 12047, United States}

\date{}

\begin{document}
\maketitle

\blfootnote{$^*$ Corresponding author (pans2@rpi.edu).}


\vspace{-.2in}
\begin{abstract}
\texttt{PyKoopman} is a Python package for the data-driven approximation of the Koopman operator associated with a dynamical system. The Koopman operator is a principled linear embedding of nonlinear dynamics and facilitates the prediction, estimation, and control of strongly nonlinear dynamics using linear systems theory. In particular, \texttt{PyKoopman} provides tools for data-driven system identification for unforced and actuated systems that build on the equation-free dynamic mode decomposition (DMD)~\cite{schmid2010jfm} and its variants~\cite{Kutz2016book,schmid2022dynamic,Brunton2022siamreview}. In this work, we provide a brief description of the mathematical underpinnings of the Koopman operator, an overview and demonstration of the features implemented in \texttt{PyKoopman} (with code examples), practical advice for users, and a list of potential extensions to \texttt{PyKoopman}.  Software is available at \url{https://github.com/dynamicslab/pykoopman}.

\vspace{0.05in}
\vspace{-.15in}
\end{abstract}

\section{Introduction}

Engineers have long relied on linearization to bridge the gap between simplified, linear descriptions where powerful analytical tools exist, and the intricate complexities of nonlinear dynamics where analytical solutions are elusive~\cite{ljung2010arc,wright1999numerical}. 
Local linearization, implemented via first-order Taylor series approximation, has been widely used in system identification~\cite{ljung2010arc}, optimization~\cite{wright1999numerical}, and many other fields to make problems tractable. 
However, many real-world systems are fundamentally nonlinear and require solutions outside of the local neighborhood where linearization is valid. 
Rapid progress in machine learning and big data methods are driving advances in the data-driven modeling of such nonlinear systems in science and engineering~\cite{Brunton2019book}
Koopman operator theory in particular has emerged as a principled approach to embed nonlinear dynamics in a linear framework that goes beyond simple linearization~\cite{Brunton2022siamreview}. 

In the diverse landscape of data-driven modeling approaches, Koopman operator theory has received considerable attention in recent years~\cite{Budivsic2012chaos,Mezic2013arfm,Williams2015jnls,klus2017data,Li2017chaos,Brunton2017natcomm}. These strategies encompass not only linear methodologies~\cite{Nelles2013book,ljung2010arc} and dynamic mode decomposition (DMD)~\cite{schmid2010jfm,rowley2009spectral,Kutz2016book}, but also more advanced techniques such as nonlinear autoregressive algorithms~\cite{Akaike1969annals,Billings2013book}, neural networks~\cite{long2017pde,yang2020physics,Wehmeyer2018jcp,Mardt2018natcomm,vlachas2018data,pathak2018model,lu2019deepxde,Raissi2019jcp,Champion2019pnas,raissi2020science}, Gaussian process regression~\cite{raissi2017parametric}, operator inference, and reduced-order modeling~\cite{Benner2015siamreview,peherstorfer2016data,qian2020lift}, among others~\cite{Giannakis2012pnas,Yair2017pnas,bongard_automated_2007,schmidt_distilling_2009,Daniels2015naturecomm,brunton2016pnas,Rudy2017sciadv}. The Koopman operator perspective is unique within data-driven modeling techniques due to its distinct aim of learning a coordinate system in which the nonlinear dynamics become linear.  This methodology enables the application of closed-form, convergence-guaranteed methods from linear system theory to general nonlinear dynamics. To fully leverage the potential of data-driven Koopman theory across a diverse range of scientific and engineering disciplines, it is critical to have a central toolkit to automate state-of-the-art Koopman operator algorithms.

\texttt{PyKoopman} is a Python package designed to approximate the Koopman operator associated with both natural and actuated dynamical systems from measurement data. Specifically, \texttt{PyKoopman} offers tools for designing observables (i.e., functions of the system state) and inferring a finite-dimensional linear operator that governs the dynamic evolution of these observables in time. These steps can either be conducted sequentially~\cite{Williams2015jcd,Williams2015jnls} or combined, as demonstrated in more recent neural network models~\cite{lusch2018deep,otto2019linearly,Mardt2018natcomm,Takeishi2017nips}. 
Once a linear embedding is discovered from the data, the linearity of the transformed dynamical system can be leveraged for enhanced interpretability~\cite{pan2021sparsity} or for designing near-optimal observers~\cite{surana2016linear} or controllers for the original nonlinear system~\cite{korda2020optimal,mauroy2020koopman,kaiser2021data,peitz2019koopman,peitz2020data}.

The \texttt{PyKoopman} package is designed for both researchers and practitioners, enabling anyone with access to data to discover embeddings of nonlinear systems where the dynamics become approximately linear. Following \texttt{PySINDy}~\cite{de2020pysindy} and \texttt{Deeptime}~\cite{hoffmann2021deeptime}, \texttt{PyKoopman} is structured to be user-friendly for those with basic knowledge of linear systems, adhering to \texttt{scikit-learn} standards, while also offering modular components for more advanced users.

\section{Background}\label{sec:background}

\texttt{PyKoopman} provides Python implementations of several leading algorithms for the data-driven approximation of the Koopman operator associated with a dynamical system 
\begin{equation}
\label{eq:controlled_dynamical_system}
\frac{d}{dt}\bx(t) = \bdyn(\bx(t), \bu(t)),
\end{equation}
where $\bx \in \mathcal{M}\subseteq\mathbb{R}^n$ is the state of the system and $\mathbf{f}$ is a vector field describing the dynamics and the effect of control input $\bu\in\mathbb{R}^q$. For the sake of simplicity, we will only present the background for the autonomous dynamical system, and more details for non-autonomous dynamical systems can be found in \cref{apdx:theory}.

Consider the autonomous system
\begin{equation}
\label{eq:dynamical_system}
\frac{d}{dt}\bx(t) = \bdyn(\bx(t)).
\end{equation}
Data are typically sampled discretely in time in intervals of $\Delta t$, and the corresponding discrete-time dynamical system is given by the nonlinear map $\mathbf{F}: \mathcal{M} \mapsto \mathcal{M}$, 
\begin{equation}
\label{eq:discrete_dynamical_system}
\bx(t+\Delta t) = \mathbf{F}(\bx(t)), 
\end{equation}
where $\mathbf{F}(\bx) = \bx(t) + \int_{t}^{t+\Delta t} \bdyn(\mathbf{x}(s))\,ds$.

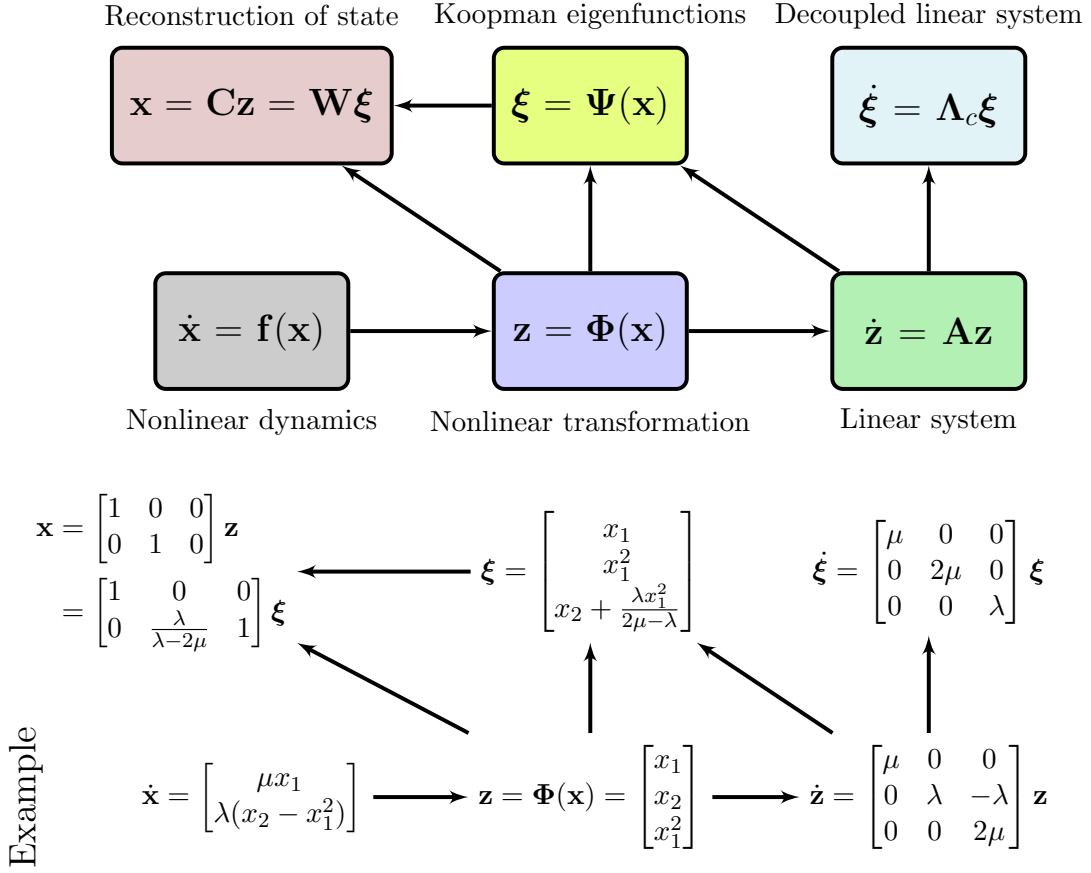
\begin{figure}
  \centering
  \begin{tikzpicture}[node distance=4cm]
    \node [block, line width=1.5pt, text width=6em, fill=black!20] (NonDyn) {\Large$\dot{\bx}=\bdyn(\bx)$};
    \node [block, line width=1.5pt, right of=NonDyn, node distance=4.5cm, text width=6em] (Lift) {\Large$\bz = \bPhi(\bx)$};
    \node [block, line width=1.5pt, right of=Lift, fill=green!80!black!30, node distance=4.5cm, text width=6em] (LinDyn) {\Large$\dot{\bz} = \bA\bz$};
    \node [block, line width=1.5pt, above of=NonDyn, fill=red!50!black!20, node distance=3cm, text width=9em] (Rec) {\Large$\bx = \bC \bz = \bW \bxi$};
    \node [block, line width=1.5pt, above of=LinDyn, fill=cyan!80!green!10, node distance=3cm, text width=6em] (DPLinDyn) {\Large$\dot{\bxi} = \bLambda_c \bxi$};
    \node [block, line width=1.5pt, above of=Lift, fill=green!20!yellow!50, node distance=3cm, text width=6em] (DPRec) {\Large$\bxi = \bPsi (\bx)$};

    \path [line, line width=1.5pt] (NonDyn) -- (Lift);
    \path [line, line width=1.5pt] (Lift) -- (LinDyn);
    \path [line, line width=1.5pt] (Lift) -- (Rec);
    \path [line, line width=1.5pt] (DPRec) -- (Rec);

    \node [below of=NonDyn, node distance=1.2cm, line width=1.5pt] (NonDynText){Nonlinear dynamics};
    \node [below of=Lift, node distance=1.2cm, line width=1.5pt] (LiftText) {Nonlinear transformation};
    \node [below of=LinDyn, node distance=1.2cm, line width=1.5pt] (LinDynText) {Linear system};
    \node [above of=Rec, node distance=1.2cm, line width=1.5pt] (RecText) {Reconstruction of state};
    \node [above of=DPRec, node distance=1.2cm, line width=1.5pt] (DPRecText) {Koopman eigenfunctions};
    \node [above of=DPLinDyn, node distance=1.2cm, line width=1.5pt] (DPLinDynText) {Decoupled linear system};
    
    \node [below of=NonDynText, node distance=5cm, line width=1.5pt] (NonDynEx) {$\dot{\bx}=\begin{bmatrix}\mu x_1\\ \lambda(x_2-x_1^2) \end{bmatrix}$};
    \node [below of=LiftText, node distance=5cm] (LiftEx) {$\bz = \bPhi(\bx) =\begin{bmatrix} x_1 \\ x_2 \\ x_1^2 \end{bmatrix}$};
    \node [above of=NonDynEx, node distance=3cm, xshift=-12mm] (RecEx) {
    $\begin{aligned}
    \bx &= 
    \begin{bmatrix} 1 & 0 & 0\\ 0 & 1 & 0 \end{bmatrix}\bz
    \\
    &=\begin{bmatrix}
        1 & 0 & 0\\
        0 & \frac{\lambda}{\lambda-2\mu} & 1
    \end{bmatrix}\bxi
    \end{aligned}$
    };
    \node [above of=LiftEx, node distance=3cm] (DPRecEx) {$\bxi =\begin{bmatrix} x_1 \\ x_1^2 \\ x_2 +  \frac{\lambda x_1^2}{2\mu-\lambda}\end{bmatrix}$};
    \node [below of=LinDynText, node distance=5cm] (LinDynEx) {$\dot\bz =\begin{bmatrix} \mu & 0 & 0\\ 0 & \lambda & -\lambda \\ 0 & 0 & 2\mu \end{bmatrix}\bz$};
    \node [above of=LinDynEx, node distance=3cm] (DPLinDynEx) {$\dot\bxi =\begin{bmatrix} \mu & 0 & 0\\ 0 & 2\mu & 0 \\ 0 & 0 & \lambda \end{bmatrix}\bxi$};
    
    \node [left of=NonDynEx, node distance=3cm] {\rotatebox{90}{\Large Example}};
    
    \path [line, line width=1.5pt] (NonDynEx) -- (LiftEx);
    \path [line, line width=1.5pt] (LiftEx) -- (LinDynEx);
    \path [line, line width=1.5pt] (LiftEx) -- (RecEx);
    \path [line, line width=1.5pt] (LiftEx) -- (DPRecEx);
    \path [line, line width=1.5pt] (LinDyn) -- (DPLinDyn);
    \path [line, line width=1.5pt] (LinDyn) -- (DPRec);
    \path [line, line width=1.5pt] (Lift) -- (DPRec);
    \path [line, line width=1.5pt] (DPRecEx) -- (RecEx);

    \path [line, line width=1.5pt] (LinDynEx) -- (DPLinDynEx);
    \path [line, line width=1.5pt] (LinDynEx) -- (DPRecEx);
    
  \end{tikzpicture}
  \caption{Lifting of the state $\bx$ of the continuous autonomous dynamical system in \cref{eq:dynamical_system} into a new coordinate system, in which the original nonlinear dynamics become linear and are easier to handle. One can also linearly reconstruct the state $\bx$ from the new coordinate system. This is facilitated with \texttt{PyKoopman} in a data-driven manner.}
  \label{fig:LinearizingTransformation}
\end{figure}
Given data in the form of measurement vectors $\bx(t)$, the goal of data-driven Koopman theory (see \cref{fig:LinearizingTransformation}) is to find a new coordinate system
\begin{equation}
	\label{eq:transform_phi}
    \bz:=\bPhi(\bx),
\end{equation}
where the dynamics are simplified, or ideally, linearized in the sense of either continuous dynamics, 
\begin{equation}
	\label{eq:lifted_continuous_dynamics}
    \frac{d}{dt}\bz = \bA_c \bz, 
\end{equation}
or discrete-time dynamics, 
\begin{equation}
	\label{eq:lifted_discrete_dynamics}
    \bz(t+\Delta t) = \bA \bz(t),
\end{equation}
where the subscript $c$ is for continuous-time and $\bA = \exp({\Delta t \bA_c })$. For simplicity, \texttt{PyKoopman} is focused on the discrete dynamical system in \cref{eq:lifted_discrete_dynamics}, which is consistent with the majority of the literature~\cite{Kutz2016book,schmid2022dynamic,Brunton2022siamreview}. 

The goal of learning the coordinates $\bPhi$ and linear dynamics $\bA$ may be posed as a regression problem in terms of finding the linear operator that best maps the state of the system, or a transformed version of the state, forward in time. This may be formulated in terms of the following two data matrices,
\begin{equation}
\label{eq:data_matrix}
\bX = \begin{bmatrix} \vline & \vline & & \vline \\
\bx(t_1) & \bx(t_2) & \cdots & \bx(t_m) \\
 \vline & \vline & & \vline
 \end{bmatrix}, 
 \quad     
 \bX' = \begin{bmatrix} \vline & \vline & & \vline \\
\bx(t_1') & \bx(t_2') & \cdots & \bx(t_m') \\
 \vline & \vline & & \vline
 \end{bmatrix},
\end{equation}
or the transformed data matrices of candidate nonlinear observations
\begin{equation}
\label{eq:transformed_data_matrix}
\bPhi(\bX) = \begin{bmatrix} \vline & \vline & & \vline \\
\bPhi(\bx(t_1)) & \bPhi(\bx(t_2)) & \cdots & \bPhi(\bx(t_m)) \\
 \vline & \vline & & \vline
 \end{bmatrix}, 
 \bPhi(\bX')
 = \begin{bmatrix} \vline & \vline & & \vline \\
\bPhi(\bx(t_1')) & \bPhi(\bx(t_2')) & \cdots & \bPhi(\bx(t_m')) \\
 \vline & \vline & & \vline
 \end{bmatrix}.
\end{equation}
The following regression is then performed to approximately solve
\begin{equation}
\label{eq:regression}
\bPhi(\bX') \approx \mathbf{A} \bPhi(\bX)
\end{equation}
for an unknown $\bA$. 
The choice of $\bPhi$ is problem dependent. Popular choices are polynomial features~\cite{Williams2015jnls}, implicit features defined by kernel functions~\cite{Williams2015jcd}, radial basis functions~\cite{Williams2015jnls}, time delay embedding~\cite{Brunton2017natcomm}, and random Fourier features~\cite{degennaro2019scalable}. 
While most early formulations of data-driven Koopman approximation rely heavily on ordinary least squares~\cite{Brunton2019book} or SVD-DMD~\cite{schmid2010jfm}, one can use any regression from the DMD community (for example, using \texttt{PyDMD}~\cite{demo2018pydmd}) to solve \cref{eq:regression}, including total least squares (tlsDMD)~\cite{hemati2017biasing}, optimized DMD (optDMD)~\cite{Askham2018siads}, etc.

Although originating in the field of fluid dynamics~\cite{rowley2009spectral,schmid2010jfm} for modal analysis~\cite{bagheri2013koopman,pan2021sparsity,Taira2017aiaa,Taira2020aiaaj}, the Koopman operator and its variants have inspired numerous ideas in the control community, such as Koopman optimal control~\cite{brunton2016koopman,kaiser2021data}, Koopman model predictive control (MPC)~\cite{korda2018linear}, Koopman reinforcement learning~\cite{weissenbacher2022koopman}, and Koopman-based observers and Kalman filters~\cite{surana2016linear}. Furthermore, the application of the Koopman operator has been extensively employed in control-oriented model identification in fields such as robotics~\cite{mamakoukas2021derivative,abraham2019active}, weather forecasting~\cite{xiong2023koopmanlab}, and time series prediction~\cite{lange2021fourier}.
However, there is currently no standard open-source implementation for approximating the Koopman operator from data. Consequently, researchers are required to develop their own versions, even though their primary interests may be in the downstream applications of the Koopman operator. This has motivated this current work to standardize the implementation of the Koopman operator by creating \texttt{PyKoopman}. This platform is designed to serve as a central hub for Koopman operator education, experimentation with various techniques, and an off-the-shelf toolkit for end-users to seamlessly integrate data-driven Koopman algorithms into their task pipelines.

\section{Features}

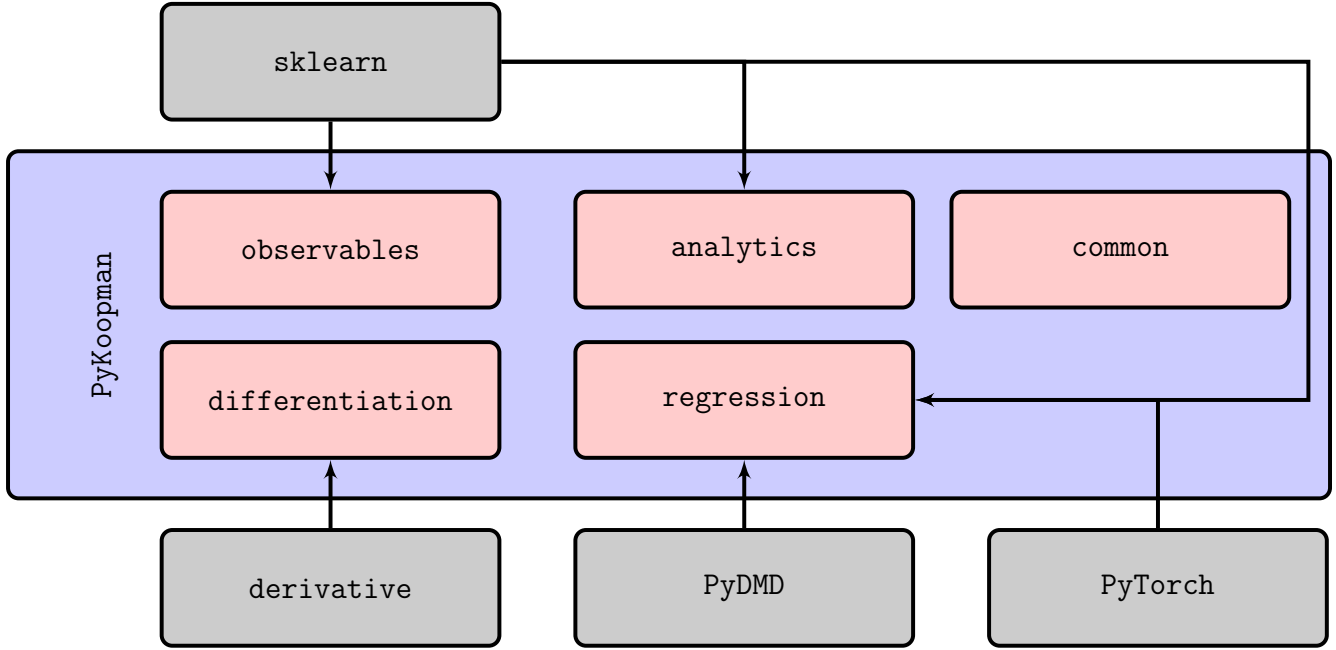
\begin{figure}
  \centering
    \begin{tikzpicture}[]
    \node[block, line width=1.5pt, fill=blue!20, text width=45em, minimum height=12em] (pykoop) {};
    \node[left of=pykoop, node distance=7.5cm, line width=1.5pt] {\rotatebox{90}{\large \texttt{PyKoopman}}};

    \node (diff) [block, below of = pykoop, line width=1.5pt, fill=red!20, text width=11em, xshift=-45mm] { \large\texttt{differentiation} };
    
    \node (obsv) [block, above of = pykoop, line width=1.5pt, fill=red!20, text width=11em, xshift=-45mm] { \large\texttt{observables} };
    
    \node (regr) [block, below of = pykoop, line width=1.5pt, fill=red!20, text width=11em, xshift=10mm] { \large\texttt{regression} };
    
    \node (anly) [block, above of = pykoop, line width=1.5pt, fill=red!20, text width=11em, xshift=10mm] { \large\texttt{analytics} };
    
    \node (comm) [block, above of = pykoop, line width=1.5pt, fill=red!20, text width=11em, xshift=60mm] { \large\texttt{common} };

    \node (deriv) [block, below of = diff, line width=1.5pt, fill=black!20, text width=11em, node distance=2.5cm] { \large\texttt{derivative} };
    
    \node (pydmd) [block, right of = deriv, line width=1.5pt, fill=black!20, text width=11em, node distance=5.5cm] { \large\texttt{PyDMD} };

    \node (pytorch) [block, right of = pydmd, line width=1.5pt, fill=black!20, text width=11em, node distance=5.5cm] { \large\texttt{PyTorch} };
    
    \node (sklearn) [block, above of = obsv, line width=1.5pt, fill=black!20, text width=11em, node distance=2.5cm] { \large\texttt{sklearn} };
    
    \draw [line, line width=1.5pt] (sklearn.east) -| (8.5,-1) -- (regr.east); 
    \path [line, line width=1.5pt] (sklearn) -| (anly);
    \path [line, line width=1.5pt] (pytorch) |- (regr);
    \path [line, line width=1.5pt] (sklearn) -- (obsv);
    \path [line, line width=1.5pt] (deriv) -- (diff);
    \path [line, line width=1.5pt] (pydmd) -- (regr);
    \end{tikzpicture}
    \caption{External package dependencies of PyKoopman.}
  \label{fig:package-structure-dependency}
\end{figure}

The core component of the \texttt{PyKoopman} package is the \texttt{Koopman} model class. To make this package accessible to a broader user base, this class is implemented as a \texttt{scikit-learn} estimator. The external package dependencies are illustrated in \cref{fig:package-structure-dependency}. Additionally, users can create sophisticated pipelines for hyperparameter tuning and model selection by integrating \texttt{pykoopman} with \texttt{scikit-learn}.

\begin{figure}
  \centering
    \begin{tikzpicture}[]
    \node[block, line width=1.5pt, fill=white!20, text width=15em, minimum height=5em] (koop) {\Large $\bz_{k+1}=\bA\bz_k$};
    \node[block, below of=koop, node distance=3cm, line width=1.5pt, fill=white!20, text width=15em, minimum height=7em] (koop_ctrl) {\Large $\begin{bmatrix}\bz\\ \bu\end{bmatrix}_{k+1}=\underbrace{\begin{bmatrix}\bA & \bB\\ \cdot & \cdot \end{bmatrix}}_{=\bK}\begin{bmatrix}\bz\\ \bu\end{bmatrix}_k$ };
    \node[left of=koop, node distance=3.5cm, line width=1.5pt] {\rotatebox{90}{\Large Unforced}};
    \node[left of=koop_ctrl, node distance=3.5cm, line width=1.5pt] {\rotatebox{90}{\Large Controlled}};
    \end{tikzpicture}
    \caption{Broad categorization of model types that can be identified with current \texttt{PyKoopman}. While the dotted parts (marked with ``$\cdot$'') can be simultaneously discovered within the framework, they are typically ignored for control purposes.}
  \label{fig:koopman-formalism}
\end{figure}
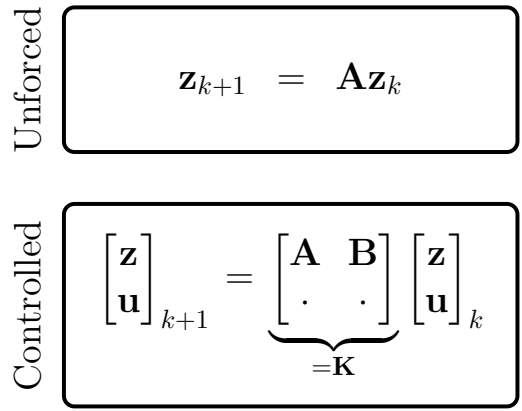

As illustrated in \cref{fig:koopman-formalism}, \texttt{PyKoopman} is designed to lift nonlinear dynamics into a linear system with linear actuation. Specifically, our \texttt{PyKoopman} implementation involves two major steps:

\begin{enumerate}
\item \texttt{observables}: the nonlinear observables used to lift $\bx$ to $\bz$, and reconstruct $\bx$ from $\bz$;
\item \texttt{regression}: the regression used to find the best-fit dynamics operator $\bA$.
\end{enumerate}

Additionally, we have a \texttt{differentiation} module that evaluates the time derivative from a trajectory and an \texttt{analytics} module for sparsifying arbitrary approximations of the Koopman operator.


At the time of writing, we have the following features implemented:
\begin{itemize}

  \item Observable library for lifting the state $\bx$ into the observable space
  \begin{itemize}
    \item Identity (for DMD/DMDc or in case users want to compute observables themselves): \texttt{Identity}
    \item Multivariate polynomials: \texttt{Polynomial}~\cite{Williams2015jnls}
    \item Time delay coordinates: \texttt{TimeDelay}~\cite{mezic2004comparison,Brunton2017natcomm}
    \item Radial basis functions: \texttt{RadialBasisFunctions}~\cite{Williams2015jnls}
    \item Random Fourier features: \texttt{RandomFourierFeatures}~\cite{degennaro2019scalable}
    \item Custom library (defined by user-supplied functions): \texttt{CustomObservables}
    \item Concatenation of observables: \texttt{ConcatObservables}
  \end{itemize}

  \item System identification method for performing regression
  \begin{itemize}
    \item Dynamic mode decomposition~\cite{schmid2010jfm,rowley2009spectral,tu2014jcd}: \texttt{PyDMDRegressor}
    \item Dynamic mode decomposition with control~\cite{proctor2016dynamic}: \texttt{DMDc}
    \item Extended dynamic mode decomposition~\cite{Williams2015jnls}: \texttt{EDMD}
    \item Extended dynamic mode decomposition with control~\cite{korda2020optimal}: \texttt{EDMDc}
    \item Kernel dynamic mode decomposition: \texttt{KDMD}~\cite{Williams2015jcd}
    \item Hankel DMD~\cite{Brunton2017natcomm}: \texttt{HDMD}
    \item Hankel DMD with control: \texttt{HDMDc}
    \item Neural Network DMD~\cite{Takeishi2017nips,lusch2018deep,Mardt2018natcomm,pan2020physics,otto2019linearly}: \texttt{NNDMD} 
  \end{itemize}
  
  	\item Sparse construction of Koopman invariant subspace
  \begin{itemize}
  	\item Multi-task learning based on linearity consistency~\cite{pan2021sparsity}: \texttt{ModesSelectionPAD21}
  \end{itemize}
  
    \item Numerical differentiation for computing $\dot{\bX}$ from $\bX$
  \begin{itemize}
    \item Finite difference: \texttt{FiniteDifference}
    \item 4th order central finite difference: \texttt{Derivative(kind=`finite\_difference')}
    \item Savitzky-Golay with cubic polynomials:
    \texttt{Derivative(kind=`savitzky-golay')}
    \item Spectral derivative:
    \texttt{Derivative(kind=`spectral')}
    \item Spline derivative:
    \texttt{Derivative(kind=`spline')}
    \item Regularized total variation derivative: \texttt{Derivative(kind=`trend\_filtered')}
  \end{itemize}

  \item Common benchmark dynamical systems
  \begin{itemize}
  \item Discrete-time random, stable, linear state-space model: \texttt{drss}
  \item Van del Pol oscillator: \texttt{vdp\_osc}
  \item Lorenz system: \texttt{lorenz}
  \item Two-dimensional linear dynamics: \texttt{Linear2Ddynamics}
  \item Linear dynamics on a torus: \texttt{torus\_dynamics}
  \item Forced Duffing Oscillator: \texttt{forced\_duffing}
  \item Cubic-quintic Ginzburg-Landau equation: \texttt{cqgle}
  \item Kuramoto-Sivashinsky equation:
  \texttt{ks}
  \item Nonlinear Schr\"{o}dinger equation: \texttt{nls}
  \item Viscous Burgers equation: \texttt{vbe}
  \end{itemize}    
  
  \item Validation routines for consistency checks
\end{itemize}

\section{Examples}

The \texttt{PyKoopman} GitHub repository\footnote{\url{https://github.com/dynamicslab/pykoopman}} provides several helpful Jupyter notebook tutorials. Here, we demonstrate the usage of the \texttt{PyKoopman} package on three low-dimensional nonlinear systems.

First, consider the dynamical system
\begin{equation}\label{eq:slow_manifold}
  \begin{aligned}
    \dot x_1 &= -0.05x_1 \\
    \dot x_2 &= -x_2 + x_1^2.
  \end{aligned}
\end{equation}

In Python, the right-hand side of \cref{eq:slow_manifold} can be expressed as follows:
\begin{lstlisting}[language=iPython]
def slow_manifold(x, t):
    return [
        -0.05 * x[0],
        -x[1] + x[0]**2
    ]
\end{lstlisting}

To prepare training data, we draw 100 random number within $[-1,1]^2$ as initial conditions and then collect the corresponding trajectories by integrating \cref{eq:slow_manifold} forward in time:
\begin{lstlisting}[language=iPython]
import numpy as np
from scipy.integrate import odeint

dt = 0.02
t = np.arange(0, 50, dt)

X = []
Xnext = []
for x0_0 in np.linspace(-1, 1, 10):
    for x0_1 in np.linspace(-1, 1, 10):
        x0 = np.array([x0_0, x0_1])
        x_tmp = odeint(slow_manifold, x0, t)
        X.append(x_tmp[:-1,:])
        Xnext.append(x_tmp[1:,:])

X = np.vstack(X)
Xnext = np.vstack(Xnext)
\end{lstlisting}
Note that \texttt{X} and \texttt{Xnext} correspond to $\bX$ and $\bX'$ in \cref{eq:data_matrix}.

\begin{figure}
     \centering
     \begin{subfigure}[b]{0.45\textwidth}
         \centering
 		 \includegraphics[width=\textwidth]{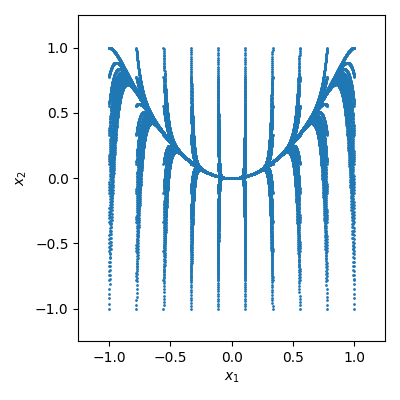}
     \end{subfigure}
     \hfill
     \begin{subfigure}[b]{0.45\textwidth}
         \centering
  		\includegraphics[width=\textwidth]{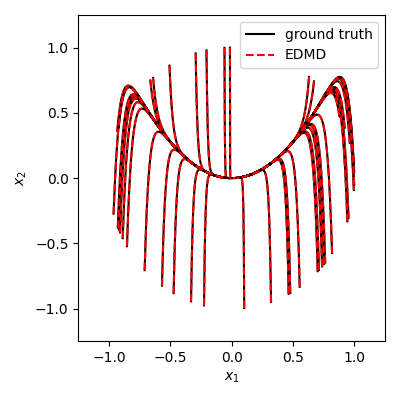}
     \end{subfigure}
     \caption{Demonstration on the slow manifold problem. \textbf{Left:} measurement data simulated using the slow manifold in \cref{eq:slow_manifold}. \textbf{Right:} Trajectories of ground truth and predictions from EDMD implemented in \texttt{PyKoopman} given unseen initial conditions.}
     \label{fig:example}
\end{figure}

We plot \texttt{X} in \cref{fig:example}, while \texttt{Xnext} is omitted for brevity. Almost all \texttt{PyKoopman} objects support this ``one-step ahead'' format of data, except when time delay is explicitly required, such as in \texttt{HAVOK}~\cite{Brunton2017natcomm}. Furthermore, \texttt{NNDMD} not only supports the standard ``one-step" ahead format but also accommodates data with multiple-step trajectories.

The \texttt{PyKoopman} package is built around the \texttt{Koopman} class, which approximates the discrete-time Koopman operator from data. To begin, we can create an observable function and an appropriate regressor. These two objects will then serve as input for the \texttt{Koopman} class. For instance, we can employ EDMD to approximate the slow manifold dynamics as shown in \cref{eq:slow_manifold}.
\begin{lstlisting}[language=iPython]
from pykoopman import Koopman
from pykoopman.observables import Polynomial
from pykoopman.regression import EDMD

model = Koopman(observables=Polynomial(2),regressor=EDMD())
model.fit(X,Xnext)
\end{lstlisting}

Once the \texttt{Koopman} object has been fit, we can use the \texttt{model.simulate} method to make predictions over an arbitrary time horizon. For example, the following code demonstrates the usage of \texttt{model.simulate} to make predictions for 50 unseen initial conditions sampled on the unit circle.

\begin{lstlisting}[language=iPython]
plt.figure(figsize=(4,4))
theta = np.random.rand(1, 50)*2*np.pi
x0_test_array = np.stack((np.cos(theta), np.sin(theta)),axis=0).T 
for x0_test in x0_test_array:
    xtest_true = odeint(slow_manifold, x0_test.flatten(), t)
    xtest_pred = model.simulate(x0_test,n_steps=t.size-1)
    xtest_pred = np.vstack([xtest_true[0], xtest_pred])

    plt.plot(xtest_true[:,0], xtest_true[:,1],'k')
    plt.plot(xtest_pred[:,0], xtest_pred[:,1],'r--')
plt.xlabel(r'$x_1$')
plt.ylabel(r'$x_2$')
\end{lstlisting}

\Cref{fig:example} displays the excellent agreement between ground truth and the EDMD prediction from the aforementioned \texttt{Koopman} model on randomly generated unseen test data. The official GitHub repository\footnote{\url{https://github.com/dynamicslab/pykoopman/tree/master/docs}} contains additional useful examples.
\section{Practical tips}\label{sec:practical-tips}

In this section, we offer practical guidance for using \texttt{PyKoopman} effectively. We discuss potential pitfalls and suggest strategies to overcome them.

\subsection{Observables selection}

The use of nonlinear observables makes the approximation of the Koopman operator fundamentally different from DMD. However, choosing observables in practice can be a highly non-trivial task. Although we used monomials as observables in the previous example, such polynomial features are not scalable for practical systems in robotics or fluid dynamics. As a rule of thumb in practice, one can try the thin-plate radial basis function~\cite{Li2017chaos} as a first choice. If the number of data snapshots in time is only a few hundred (e.g., as in fluid dynamics), one can opt for kernel DMD~\cite{pan2021sparsity}, but tuning the hyperparameters within the kernel function can be critical. If the number of data points exceeds a few thousand (e.g., multiple trajectories of simulated robotic systems), one can choose to approximate the kernel method with random Fourier features in \texttt{observables.RandomFourierFeatures} as observables~\cite{degennaro2019scalable}.

Another useful approach is time-delay observables~\cite{Brunton2017natcomm}, which can be interpreted as using the reverse flow map function recursively as observables. However, it does not self-start. Just like autoregressive models, the number of delays determines the maximum number of linearly superposable modes that the model can capture. The number of delays also has a somewhat surprising effect on the numerical condition~\cite{pan2020structure}.

Furthermore, one may find it beneficial to use customized observables informed by the governing equation in \cref{eq:dynamical_system}~\cite{ng2023data} by calling \texttt{observables.CustomObservables} with \texttt{lambda} functions. If all the above methods fail, one may choose to use a neural network to search for the observables; this approach is typically more expressive but is also more computationally expensive.

\subsection{Optimization}

Once the observables are chosen, the optimization step finds the best-fit linear operator that maps observable at the current time step to the next time step. Although most of the time the standard least-squares regression or pseudo-inverse is sufficient, one can use any regressor from \texttt{PyDMD}. Additionally, one can use \texttt{NNDMD} to concurrently search for the observables and optimal linear fit.

Regarding \texttt{NNDMD}, we have found that using the recurrent loss leads to more accurate and robust model performance than the standard one-step loss, which is adopted in more traditional algorithms. Thanks to the dynamic graph in \texttt{PyTorch}, \texttt{NNDMD} can minimize the recurrent loss progressively, starting from minimizing only the one future step loss to multiple steps in the future. Moreover, we have found that using second-order optimization algorithms, such as L-BFGS~\cite{nocedal1999numerical}, significantly accelerates training compared to the Adam optimizer~\cite{kingma2014adam}. However, occasionally the standard L-BFGS can diverge, especially when trained over a long period of time. With \texttt{PyTorch.Lightning}, \texttt{NNDMD} can easily take advantage of the computing power of various hardware platforms.

\section{Extensions}
  In this section, we list potential extensions and enhancements to our \texttt{PyKoopman} implementation. We provide references for the improvements that are inspired by previously conducted research and the rationale behind the other potential changes.

\begin{itemize}
\item \textbf{Bilinearization:} Although ideally we would like to have a standard linear input-output system in the transformed coordinates, this can lead to inconsistencies with the original system. A number of studies~\cite{qian2020lift,bruder2021advantages,goswami2021bilinearization} have shown the advantages of using bilinearization instead of standard linearization. It is worth noting that bilinearization has been incorporated into another Python package, \texttt{pykoop}~\cite{dahdah_pykoop_2022}.
\item \textbf{Continuous spectrum:} Most existing algorithms assume a discrete, pointwise spectrum reflected in the data. As a result, these algorithms may struggle with chaotic systems, which contain a continuous spectrum. There are several approaches for handling continuous spectra, including the use of time delay coordinates~\cite{Brunton2017natcomm}. Recent approaches including resDMD, MP-EDMD, and physics informed DMD all show promise for continuous-spectrum dynamics~\cite{colbrook2022mpedmd,colbrook2023residual,baddoo2023physics}.
\item \textbf{Extended libraries:} The linear system identified in the lifted space can be further exploited to facilitate the design of optimal control for nonlinear systems. For example, the classic LQR has been extended to nonlinear systems~\cite{kaiser2021data}. Moreover, nonlinear MPC can be converted to linear MPC using the identified linear system from the Koopman operator, which transforms the original non-convex optimization problem into a convex optimization problem. In the future, we believe open-source libraries for Koopman-based control synthesis integrated with \texttt{PyKoopman} will be widely used by the community.
\end{itemize}

\section{Acknowledgments}
The authors would like to acknowledge support from the National Science Foundation AI Institute in Dynamic Systems (Grant No. 2112085) and the Army Research Office ({W911NF-17-1-0306} and W911NF-19-1-0045).

\appendix
\section{Koopman operator theory}
\label{apdx:theory}

In this section, we will briefly describe Koopman operator theory for dynamical systems~\cite{Brunton2022siamreview}. Specifically, the theory for autonomous dynamical systems is presented in \cref{apdx:auto} while the theory for controlled systems is presented in \cref{apdx:control}. 

\subsection{Koopman operator theory for dynamical systems}
\label{apdx:auto}

Given the following continuous-time dynamical system, 
\begin{equation}
  \frac{d}{dt}\bx(t) = \bdyn(\bx(t)),
\end{equation}
the flow map operator, or time-\(t\) map,  \(\flow^{t} : \mathcal{M} \to \mathcal{M}\) maps initial conditions $\bx(0)$ to points on the trajectory $t$ time units in the future, so that trajectories evolve according to
\( \bx(t) = \flow^{t}( \bx(0) ) \).

The Koopman operator $\mathcal{K}^t: \mathcal{G}(\mathcal{M}) \mapsto \mathcal{G}(\mathcal{M})$ maps the measurement function $g\in\mathcal{G}(\mathcal{M})$ evaluated at a point $\bx(t_0)$ to the same measurement function evaluated at a point $\bx(t_0+t)$:
\begin{equation}
\Koop^{t} g (\bx)  = g (\flow^{t}(\bx))  \label{eq:koopman},
\end{equation}
where \(\mathcal{G}(\mathcal{M})\) is a set of \emph{measurement functions}
\(g : \mathcal{M} \to \mathbb{C}\).

The infinitesimal generator~\(\LieOp\) of the time-$t$ Koopman operator is known as the Lie operator~\cite{koopman1931pnas}, as it is the Lie derivative of \(g\) along the vector field \(\mathbf{f}(\bx)\) when the dynamics is given by \cref{eq:dynamical_system}.
This follows from applying the chain rule to the time derivative of $g(\bx)$:
\begin{align}
\frac{d}{dt} g(\bx(t)) &= \nabla g \cdot \dot{\bx} (t) =  \nabla g \cdot \mathbf{f}( \bx(t) )  = \LieOp g(\bx(t)). \label{eq:Lie-operator}
\end{align}

In continuous-time, a Lie operator eigenfunction \(\varphi(\bx)\) satisfies
\begin{equation}
\frac{d}{dt}\varphi(\bx) = \LieOp\varphi(\bx) = \mu \varphi(\bx).\label{Eq:KoopmanEfun}
\end{equation}
An eigenfunction \(\varphi\)  of \(\LieOp\) with eigenvalue \(\mu\) is then an eigenfunction of \(\Koop^{t}\) with eigenvalue \({\lambda^{t} = \exp(\mu t)}\).
However, we often take multiple measurements of a system, which we will arrange in a vector $\bg$:
\begin{align}
\bg(\bx) = \begin{bmatrix} g_1(\bx)\\ g_2(\bx)\\ \vdots \\ g_p(\bx)\end{bmatrix}.\label{Eq:Koopman:Measurement}
\end{align}
The vector of observables, $\bg$, can be expanded in terms of a basis of eigenfunctions $\varphi_j(\bx)$:
\begin{align}
\Koop^t \bg(\bx) =  \sum_{j=1}^{\infty}\lambda_j^t\varphi_j(\bx) \bv_j,\label{Eq:KoopmanMode}
\end{align}
where $\bv_j:=[\langle \varphi_j,g_1 \rangle,\langle \varphi_j,g_2 \rangle,\ldots,\langle \varphi_j,g_p \rangle]$ is the $j$-th \emph{Koopman mode} associated with the eigenfunction $\varphi_j$.

For a discrete-time system
\begin{eqnarray}
\mathbf{x}_{k+1}=\flow(\mathbf{x}_k),\label{Eq:DiscreteDynamics}
\end{eqnarray}
where $\bx_k=\bx(t_k)=\bx(k\Delta t)$, the Koopman operator $\mathcal{K}$ governs the one-step evolution of the measurement function $g$, 
\begin{eqnarray}
\Koop g(\bx_k) = g(\flow(\bx_k)) = g(\bx_{k+1}).
\end{eqnarray}
In this case, a Koopman eigenfunction \(\varphi(\bx)\) corresponding to an eigenvalue \(\lambda\) satisfies
\begin{equation}
\varphi(\bx_{k+1}) = \Koop \varphi(\bx_k) = \lambda\varphi(\bx_k).\label{Eq:KoopmanEfun:Discrete}
\end{equation}
%

\subsection{Koopman theory for controlled systems}
\label{apdx:control}

The continuous-time dynamics for a controlled system is given by 

\begin{equation}\label{eq:controlled_dynamical_system}
  \frac{d}{dt}\bx(t) = \bdyn(\bx(t),\bu(t)). 
\end{equation}

Following Proctor et al.~\cite{Proctor2017siads} and Kaiser et al.~\cite{kaiser2021data}, instead of the usual state $\bx$, we consider measurement functions defined on an extended state $\tilde{\bx} = (\bx,\bu)$, where the corresponding flow map
is $\tilde{\bF}^t(\bx,\bu) = [\bF^t(\bx,\bu),\bTheta^t(\bu)]$, and $\bTheta^t(\bu)$ is the shift map by time $t$ units so that $\bTheta^t(\bu)(s) = \bu(s+t)$.

In summary, the Koopman operator on controlled system governs the measurement function of the extended state,
\begin{equation}
\Koop^t g(\bx,\bu) = g(\tilde{\bF}^t(\bx,\bu)).
\end{equation}

The corresponding Koopman mode decomposition for a vector of observables, 
\begin{align}
\bg(\bx,\bu) = \begin{bmatrix} g_1(\bx,\bu)\\ g_2(\bx,\bu)\\ \vdots \\ g_p(\bx,\bu)\end{bmatrix},
\label{Eq:Koopman:Measurement}
\end{align}
can be written as,
\begin{align}
\Koop^t \bg(\bx,\bu) =  \sum_{j=1}^{\infty}\lambda_j^t\varphi_j(\bx,\bu) \bv_j,\label{Eq:KoopmanModeControl}
\end{align}
where the Koopman eigenfunction is
\begin{equation}
    \varphi(\bx,\bu,t) = \Koop^t \varphi(\bx,\bu) = \lambda\varphi(\bx,\bu).
\end{equation}
%


If the continuous-time controlled system is control-affine,
\begin{equation}
\bdyn(\bx(t),\bu(t)) = \bdyn_0 (\bx) + \sum_{i=1}^q \bdyn_i (\bx) u_i,
\end{equation}
where $u_i$ is $i$th component of input $\bu$, then the Lie operator (along the vector field $\bdyn$) on the measurement function $g(\bx)$ becomes,
\begin{equation}
    \LieOp g(\bx) 
    = \nabla_{\bx} g(\bx)\cdot \dot{\bx} 
    = \nabla_{\bx} g(\bx)\cdot \bdyn_0(\bx) + 
      \nabla_{\bx} g(\bx)\cdot \sum\limits_{i=1}^q \bdyn_i(\bx)u_i.
\end{equation}

Similarly, after we define the Lie operator along the vector field $\bdyn_0$ as $\mathcal{A}$ and that along $\bdyn_i$ as $\mathcal{B}_i$, we have the bilinearization for the control-affine system, 
\begin{equation}
    \ddt{g}(\bx) = \mathcal{A}g(\bx) + \sum\limits_{i=1}^q u_i \mathcal{B}_ig(\bx).
\end{equation}

Assuming $\varphi$ is an eigenfunction of $\mathcal{A}$, we have
\begin{equation}
    \ddt{\varphi}(\bx) 
    = \mu \varphi(\bx) + 
      \nabla_{\bx} \varphi(\bx)\cdot \sum\limits_{i=1}^q \bdyn_i(\bx)u_i.
      \label{eq:koopman_for_control_affine}
\end{equation}

Furthermore, if the vector space spanned by $D$ such eigenfunctions $\{\varphi_i\}_{i=1}^{D}$ is invariant under $\mathcal{B}_1,\ldots,\mathcal{B}_q$~\cite{otto2021koopman}, we have
\begin{equation}
    \forall i = 1,\ldots,q, \quad \mathcal{B}_i \bm{\varphi} = \mathbf{B}_i \bm{\varphi}, 
\end{equation}
where $\bm{\varphi} = \begin{bmatrix} \varphi_1 & \ldots & \varphi_D \end{bmatrix}^\top$.

Plugging this into \cref{eq:koopman_for_control_affine}, we have the well-known \emph{Koopman bilinear form} for the control-affine systems, 
\begin{equation}
    \ddt{\bm{\varphi}}(\bx) 
    = \bLambda_c \bm{\varphi}(\bx) + 
      \sum\limits_{i=1}^q u_i \bB_i \bm{\varphi}  .
\end{equation}
%


For general discrete-time system, 
\begin{eqnarray}
\bx_{k+1}=\flow(\bx_k,\bu_k),\label{Eq:DiscreteDynamics}
\end{eqnarray}
where $\bx_k=\bx(t_k)=\bx(k\Delta t)$, the Koopman operator governs the one-step evolution of the measurement function $g$ of the extended state $\tilde{x} = (\bx,\bu)$,
\begin{eqnarray}
\Koop g(\bx_k,\bu_k) = g(\flow(\bx_k,\bu_k)) = g(\bx_{k+1},\bu_{k+1}).
\end{eqnarray}
A Koopman eigenfunction \(\varphi(\bx)\) corresponding to an eigenvalue \(\lambda\) satisfies
\begin{equation}
\varphi(\bx_{k+1},\bu_{k+1}) = \Koop \varphi(\bx_k,\bu_k) = \lambda\varphi(\bx_k,\bu_k).\label{Eq:KoopmanEfun:Discrete}
\end{equation}
%



\newpage
\begin{spacing}{.9}
  \small{
    \setlength{\bibsep}{6.5pt}
\bibliographystyle{IEEEtran}
    \bibliography{references}
  }
\end{spacing}

\end{document}